\documentstyle[amsfonts,aps]{revtex}

\twocolumn
\begin{document}


\title{More on ``Atomic jumps in quasiperiodic Al$_{72.6}$Ni$_{10.5}$Co$_{16.9}$
and related crystalline material''} 
\author{Gerrit Coddens}

\address{Laboratoire des Solides Irradi\'es, Ecole Polytechnique,
F-91128-Palaiseau CEDEX, France}   
\date{today}   

\maketitle

\widetext 
\begin{abstract}
We strongly disagree with a number of statements  
by Dolin{\v{s}}ek et al. about 
phason dynamics in quasicrystals (QCs).\\
\end{abstract}

\narrowtext 

We are shocked by the methods used in 
the Reply of Dolin{\v{s}}ek and Dubois \cite{Reply} to 
a Comment \cite{Comment} of ours
to their paper,\cite{PRB}
which unduly  thwart our attempts of clarification.

(1) We would like to draw the attention
of the reader to a whole series of repeated
attempts to put two completely disparate
phenomena under a common denominator (1a-c).
They serve as a platform to qualify
phason dynamics 
as not special.
 
(1a) Phason dynamics and regular atomic jumps in a B2-based phase
would be the same physics since they are both leading to a broad
(fast) quasielastic neutron scattering signal.\cite{Dahlborg}

(1b) Phason dynamics and regular atomic jumps in a B2-based
phase would be the same physics since
phason sites in QCs and structural vacancies in
B2-based phases would be
both "structural vacancies" according to an akward new definition,
introduced {\em ad hoc} by the authors.\cite{Reply}

(1c) Phason dynamics and regular atomic jumps in a B2-based phase
would be the same physics since
both QCs and B2-based phases would be "open structures".\cite{Reply}
In reality, there is no proof for the claim 
that a QC would be an "open structure".\cite{illdefined}
The authors ``derive'' it {\em inductively} from the wrong premises (1a)-(1b).
But the fact that a number of sites are offered to an atom
to jump to, does not forcedly imply that the structure is ``open'',
as the counterexample of interstitial sites
clearly shows. (Furthermore, the fact that phason jumps in quasicrystals
are assisted means that there is more to it than
just the presence of a site available. Certain atoms might
have to move temporarily out of the way to render the jump possible). 

What characterizes if something is special,
are the features that it has {\em not} in common with other
things. Drawing 
a long list of indistinctive features shared with
other things is not a valid procedure to prove 
that something is not special.
Such an argument is based on an invalid inversion and/or negation
of the logical implication.\cite{platypus} 

 The introduction of claim (1c) is a remarkable construction,
that is based on the same leitmotiv of incorrect inversion and/or negation 
of the logical implication. It is easily recognized that claims (1a-b) are harsh and wrong.
With claim (1c) the authors take us
upwards an {\em ad hoc} deductive (implication) scheme, 
towards a less outspoken situation.
It is much more difficult to discuss the claim (1c) and 
to prove conclusively that it is wrong
on the ba-\\

\vspace{0.75cm}

\noindent sis of presently 
available data than claims (1a-b).\cite{holes} 
The difficulty to come to terms with claim (1c)  must permit
doubts to linger on about the issues (1a)-(1b), by suggesting
 that they would follow rather logically from (1c), and that
 (1c) is plausible.
In reality, as we mentioned, (1c) 
has been artificially postulated to yield (1a)-(1b): 
It is thus (1c) that has been ``derived'' from  (1a)-(1b)
rather than (1a)-(1b) from (1c).
But we are not obliged to discuss the truth of
postulate (1c),
as it is totally
obvious that (1a)-(1b) are wrong,
independently of the issue if postulate (1c) is wrong or otherwise.
It is also totally obvious
that phason dynamics in a QC and regular atomic jumps in a B2-based
phase are not the same physics, even if the signals they give rise to
may look similar in certain selected aspects.

(2) In the repeated attempts to identify phason dynamics with conventional
atomic hopping in a B2-based phase,  each time contradicting
evidence has been ignored. 

(2a) Making use of SrTiO$_{3}$ to depicture phason jumps
in QC as nothing special is not appropriate because
the corresponding high-temperature oxygen dynamics in 
SrTiO$_{3}$ are not ultrafast
as in QCs. 

(2b) Similarly,
in the B2-based phase, the jumps are indeed fast like in QCs, 
but here it is the geometry, temperature dependence, etc... that do not blend
into the picture. The jump distances are not short as in QCs
and instead the fast dynamics are a trivial consequence
of a huge vacancy concentration, which has no counterpart in QCs.

In both cases
it is tacitly denied that there is more to phason dynamics than just
the one single aspect that is being picked. Phason dynamics are
fast, there is a special geometry involved in the phason site
that is an intrinsic part of the definition of the quasilattice itself,
phasons have
an unusual, special temperature dependence, 
there could exist correlated simultaneous jumps
of several atoms, etc...

(3) The authors present the reader with a long, technical
demonstration that the signal in AlCuNi is not magnetic. The motivation
for this should be that
I would have claimed 
that the NMR signal in AlNiCo is due
to magnetism. The reader can verify in the text of my Comment that
I never stated that the signal in AlNiCo is magnetic. 
I addressed AlPdMn  which is well known to be magnetic 
for certain compositions,
a fact the authors admit themselves.
I clearly stated that I wanted to mention
magnetism as a possibility {\em in principle}, in order to show
that the interpretation of NMR data is not as straightforward
as the authors try to make us believe (by not providing
any discussion at all), and that the authors
have already been misguided by their lack of rigor in the past,
presenting a magnetic signal as evidence for phason dynamics.

(4) The purpose of my Comment was  to 
indicate that the interpretation of the authors is wrong
and that the authors have not bothered about
eliminating alternative, more reasonable interpretations
of their data.
If my preference goes to
an interpretation of their data in the form
of some small non-phasonic atomic displacements, due
to the presence of lattice distortions induced 
by structural defects, I pointed out
that this does not mean
that I claim any interpretation at all.
The burden of the (perhaps presently inaccessible) 
correct interpretation
and its proof remains entirely with the authors,
and one cannot 
reverse this situation by
picking one example in my objection that there are many
other possible interpretations {\em in principle}, 
in order to mispresent  my Comment as a wrong interpretation of the data.
That the authors go into great efforts to explain that the signal
in AlNiCo is not magnetic is perhaps fine, but not really needed.
It distracts the attention of the reader from the essence
of the discussion and
it is already well established that AlNiCo is not magnetic.

(5) As the authors now admit, the attribution of the NMR signal
in reference \cite{Duboisfaux} was wrong.
At the time I pointed out in a Comment \cite{CommentDubois} 
that the temperature
behavior of their signal could not possibly
agree with their interpretation in terms of phason dynamics.
The Reply \cite{ReplyDubois} thwarted the recognition
of the fact
 that that conclusion was final.
The authors postponed admitting that the signal was due to magnetism
to another occasion.\cite{PhysB}
Nonetheless, in reference \cite{PRB} the authors claimed again that
the signal observed in reference \cite{Duboisfaux,crapreview}
is a valid observation of phason dynamics and the Kalugin-Katz transition.
And in the present reply
they admit once again that this is wrong and that the AlPdMn data have 
been overinterpreted.

(6) We would like to draw the reader's attention to the use
the authors make of a paper by Yan et al. \cite{Pennycook}
who erroneously called phason sites vacancies.
This was pointed out in a Comment \cite{CommentPennycook} but 
 clarification was thwarted by a statement in the Reply by Yan et al.
 \cite{ReplyPennycook}
that they used a different definition 
of the compound word "columnar vacancy",
in order to escape from the criticism that
they introduced a different definition of the single word
"vacancy". This non-conventional 
definition had not been given
in the original paper and was just introduced {\em a posteriori} 
in the Reply.
Such {\em ad hoc} methods are not viable and can only
produce  confusion.
Dolinsek and Dubois built on  that confusion twice:
(a) They claimed that AlNiCo contains a lot of "vacant sites"
citing the paper by Yan et al., ignoring my Comment
on it. (b) Now I have pointed out once again that
phason sites are not vacancies, they take the same 
loophole of escape as Yan et al.
and state that they use a different definition
for the compound word "structural vacancy". Again, the use of a different 
definition is explicited a posteriori, and at variance with
normal, legitimate expectations.
The scientific convenience of such a definition 
has been discussed under (1).

(7) The authors say that they have been cautious by stating that the data
are {\em compatible} with phason dynamics. This is perhaps not a direct claim, 
but nevertheless far too much of a claim. The caution does not serve
to warn the reader but to thwart due criticism
for trying to pass an unwarranted claim. 
Real caution would consist in stating that the data 
are also compatible with a plethora of
(more reasonable) explanations. And there are enough other arguments in my Comment
that show that the data do {\em not} correspond to phason dynamics.

(8) When I stated it is not proved that tunneling states are
not phasons, I thought that perhaps, it might 
never be possible
to prove that they are not phasons, due to the limitations of 
the techniques. This certainly does not imply that it would 
be reasonable to claim 
that tunneling states are phasons:  There is more in being right than not
being proved wrong. 
In fact, the authors do not mention that there is even less
proof that tunneling states would be phasons. No argument is given
to substantiate their claim, except the experimental difficulties 
one encounters to provide the 
absolute final proof of the contrary.
The reader will certainly be surprized to learn that 
in reference \cite{archiv} the authors 
state themselves
that ``tunneling are definitely not an intrinsic property
of quasicrystals'', contradicting their own claims! 

Let us add to that that the double wells
in phason dynmaics are defined on Euclidean three-dimensional space,
while those in tunneling states are defined on an abstract configuration
space. The {\em new} interpretation proposed by Dolinsek and Dubois rests thus
on a confusion between  two very different realities covered 
by the same words "double well". 
We pointed out that phason dynamics in QCs explore a discrete set
of jump vectors, in contrast to what is considered to happen in tunneling
states. 
That was already an excellent indication that
tunneling states are not phasons.

(9) That  
it is not possible to 
link the high-temperature neutron
signals to their NMR data, cannot justify
the high-profile claims of the authors. 
They try to save their interpretation
of the NMR signals in an amended version by
claiming that there could be
a cross-over of dynamical regime from hopping
to tunneling. The surprizing character
of this claim should not disarm or confuse the reader:
It
cannot make away with our arguments based on the X-ray data. 
We  pointed out that in AlNiCu there is clear evidence
from X-ray diffraction
for an order-disorder transition. Below the 
transition temperature,
the jump dynamics are just frozen. 
X-ray diffraction would not give evidence for 
an order-disorder transition
if the atomic motion just crossed over 
from over-the-barrier hopping
to tunneling, because it would still detect the 
instantaneous positional disorder, 
created by the atomic motion.

Furthermore, it must be clear
that the distances in AlNiCu are  too large
(and the transition temperature too high)
to consider tunneling 
of the type proposed by the authors
as a real possibility.
A comparison with tunneling in SrTiO$_{3}$ is 
thus totally inappropriate.

On the same footing, it is generally admitted
that the freezing of phason hopping
in QCs is responsible for the increase of diffuse scattering
observed below a certain temperature (which is e.g. $\approx$ 600 C
in AlPdMn). If there were a cross-over from
phason hopping to phason tunneling, there would
be no change in the X-ray data below the
cross-over temperature.

(10) I already pointed out that in AlPdMn 
the assistance energies
are not close to 0.6 eV. Until much more 
serious proof  is given 
caution requires thus to consider 
the numerical similarities as coincidences.

(11) The authors deny me the use of
the word "phason", although I come first in 
matters of having spelled out
numerous times the limitations of this 
terminology.\cite{Europhys,3ax} 
Moreover,
the latter criticism could be extended to 
the whole literature on QCs - often on much more 
serious grounds \cite{deBoissieu} -
rather than just focusing it onto the subject 
of my work on phason dynamics, the
significance of which extends far beyond
a mere matter of terminology.

(12) The farfetched, invalid  
criticisms of this group on my work
\cite{PRBlistattacks} make a long, damaging, disparate list. 
The motivations given for the present attacks contradict each other.
 The authors state to have 
selected {\em on purpose} a sample with an exceptional amount of structural
vacancies  to permit them to qualify phason dynamics  
as ``not special'' \cite{Dahlborg} or ``not QC-specific''.
But previously  they have stated to be 
``surprized'' 
as they ``did {\em not intend} to question 
the excellent work on phason dynamics''.\cite{Dahlborgreply}
It is thus just that all this excellent work has been witlessly misaimed at
forcing an open door. The tacit implications of this declaration of intention 
of the authors carry strong personal undertones. The same can be said
about the gratuitous, subjective rating that phason dynamics
are not special.
But both issues
can even not be touched upon
without attracting further personal discredit.


\begin{thebibliography}{99}
\bibitem{illdefined} The terminology ``open structure''
is vague and not clearly defined.
\bibitem{platypus} We can illustrate the relevance of these methods 
by an analogy. The platypus
is a unique animal in that is both a mammal and laying eggs.
The argument of the authors is analogous to
claiming that there is  nothing special about the platypus
as birds are laying eggs as well.
\bibitem{holes} This is due to the lack
of adequate experimental procedures: QC single-domain samples contain large
holes that thwart a reliable determination of their density.
\end{thebibliography}
\end{document}